\documentclass{Interspeech2024}

\interspeechcameraready 

\usepackage[dvipsnames]{xcolor}
\usepackage{multirow}
\usepackage{tikz}
\usepackage{fontawesome}
\usepackage{xspace}
\usepackage{tabularx}
\usepackage{graphbox}
\usepackage{microtype}
\usepackage{cite}

\usetikzlibrary{backgrounds} 
\usetikzlibrary{fit} 

\newcommand\etal{\textit{et al.}\xspace}
\newcommand\eg{\textit{e.g.}\xspace}

\title{Translating speech with just images}

\name[affiliation={1}]{Dan}{Oneata}
\name[affiliation={2}]{Herman}{Kamper}
\email{}

\address{
  $^1$\textsc{Politehnica} Bucharest, Romania\\
  $^2$Stellenbosch University, South Africa
}

\usepackage[prependcaption,textsize=scriptsize]{todonotes}
\definecolor{mycolor}{HTML}{FF6600}

\usepackage[prependcaption,textsize=scriptsize]{todonotes}

\begin{document}

\maketitle

\begin{abstract}
Visually grounded speech models link speech to images.
We extend this connection by linking images to text via an existing image captioning system,
and as a result gain the ability to map speech audio directly to text.
This approach can be used for speech translation with just images by having the audio in a different language from the generated captions.
We investigate such a system on a real low-resource language, Yorùbá,
and propose a Yorùbá-to-English speech translation model that leverages pretrained components in order to be able to learn in the low-resource regime.
To limit overfitting, we find that it is essential to use a decoding scheme that produces diverse image captions for training.
Results show that the predicted translations capture the main semantics of the spoken audio, albeit in a simpler and shorter form. 
\end{abstract}

\section{Introduction}

Imagine you are a linguist tasked with translating a foreign low-resource language, but that it is not possible to get parallel speech--translations.  %
One possible approach is to ask native speakers to describe images using their own language.
The idea would be to then use the images as an intermediate modality to understand new input speech~\cite{harwath2018icassp,zhao2022naacl}.
While there has been major advances in visually grounded speech models that learn from paired audio--image correspondences~\cite{harwath2020iclr, scharenborg2020taslp, sanabria2021interspeech, rouditchenko2021interspeech, chrupala2022jair}, %
no study has attempted to develop a model that can take speech and directly produce a written translation of the input.
This is our goal.

Earlier work~\cite{kamper2018sltu} has shown that it is possible to perform keyword detection in a foreign language using only images paired with unlabelled speech.
The idea was to use a pretrained vision system to tag images with word labels in the high-resource target language.
These tags were then used as targets to train an audio-to-keyword model, taking speech input in the foreign low-resource language.
At test time, the audio model was then able to predict whether a keyword (in the target language) occurred in the audio stream (of the low-resource source language).
However, the model did not predict full translations of the speech input.
Moreover, the study was done in an artificial setting where German was the high-resource target language (of the image tagger) and English audio was the low-resource source language.

An alternative approach is to do translation by retrieval: finding relevant existing captions for a given audio in a foreign language~\cite{harwath2018icassp,ohishi2020icassp,berry2023icassp}.
These methods project audio and images in a common embedding space.
Then at test time they can map a novel audio to the caption of the closest image, thereby producing a full natural language translation.
However, retrieval is limited to the dataset and requires manually provided captions.

\begin{figure}
    \centering
    \begin{tikzpicture}[
            font=\scriptsize,
            block/.style={text width=1.25cm, align=center, minimum height=0.75cm, rounded corners},
        ]
        \newcommand{\mylabel}[1]{{\color{black!50}\tiny\textsf{#1}}}

        \matrix [row sep=0.0cm, column sep=0.80cm] {
            \node[label=\mylabel{audio (Yorùbá)}] (audio) {\includegraphics[height=0.8cm, trim={0 0.5cm 0 0.5cm}, clip]{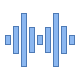}}; &
            \node[block, fill=red!20, label={\faUnlock}] (enc-audio) {audio-to-text model}; &
            \node[block, label=\mylabel{text (English)}, text width=2.0cm]  (text) {\it There are two people riding motorcycles on the road in the daytime.}; \\
            \node[label=\mylabel{image}] (image) {\includegraphics[height=1.2cm, trim={2.0cm 0.5cm 2.0cm 0.5cm}, clip]{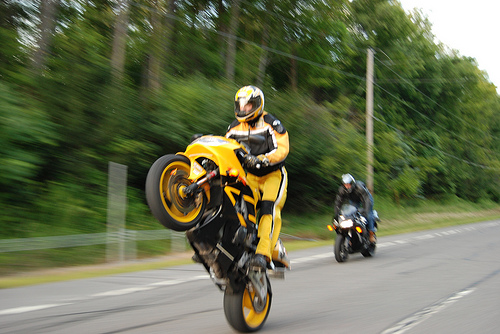}}; &
            \node[block, fill=blue!20, label={\faLock}] (enc-image) {image captioner}; & \\
        };

        \draw[->] (audio) -- (enc-audio);
        \draw[->] (image) -- (enc-image);
        \draw[->] (enc-image) -| (text);
        \draw[->] (enc-audio) -- (text);
    \end{tikzpicture}
    \caption{%
        Overview of our speech translation system.
        Given an audio in a foreign language (\eg, Yorùbá), we generate natural language translations in a high-resource language (\eg, English).
        We achieve this with only audio--image pairs by generating captions automatically using a pretrained image captioner and then using these as targets for an audio-to-text model.
    }
    \label{fig:overview}
\end{figure}
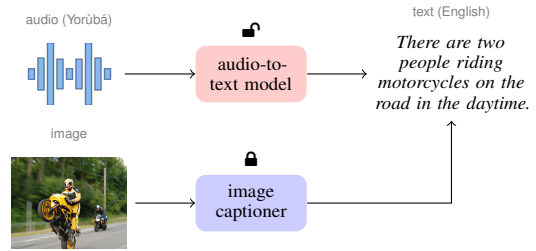

In this paper we propose a system that is able to directly generate natural language translations for a given foreign input audio.
Our speech translation system is trained solely on audio--image pairs.
The approach is illustrated in Figure \ref{fig:overview}.
First, target sentences in the high-resource language (English) are generated with a pretrained image captioning system for the image associated to an audio input.
Then, based on these sentences we learn an audio-to-text model, which takes as input speech in the foreign language (Yorùbá, in our case).
Finally, at test time we can generate translations using the audio-to-text model, in our case translating Yorùbá speech to English text.
This is done without any parallel Yorùbá--English speech--translation pairs.

In this real Yorùbá--English low-resource setting, we show that using images as an interlingua comes close to a speech translation system trained with speech--text translation pairs.
In our analysis, we also show that the same system can be used in an English--English audio-to-text system that produces reasonable paraphrases of the English audio input (again using images as intermediate modality).
By situating our results in terms of three toplines,
we conclude that it is neither the image captioning component nor the audio-to-text architecture that limits the performance;
rather, other methodological changes may be required to close the gap to human-level performance.

\section{Related work}

Our approach is an example of cross-modal learning (also referred to as cross-modal knowledge distillation).
This type of learning is applied to transfer knowledge across different modalities, for example,
from vision to depth data \cite{gupta2016cvpr} or from vision to radio signals \cite{zhao2018cvpr}.
In terms of vision and audio---the modalities of interest here---%
Ayatar \etal \cite{aytar2016neurips} used visual information to perform scene detection on audio, while
Owens \etal \cite{owens2018eccv} used ambient sound to learn visual scene information.
The work of Kamper \etal \cite{kamper2018sltu,kamper2019taslp} is more similar to our approach since it works on \textit{speech} audio.
But, as previously mentioned, they only transfer unstructured information (image tags for a fixed number of classes) rather than trying to capture the richness of natural language.
In this direction, Kim and Rush \cite{kim2016emnlp} transfer natural text by distilling the output of large translation models to smaller ones; we differ by working across modalities.

Recently, the community has also explored aligning audio features to CLIP's \cite{radford2021clip} visual features using audio--image pairs \cite{zhao2022naacl, shih2022slt}.
This approach allows to implicitly align the audio to a text embedding (via the visual channel), since CLIP provides by default a visual--text alignment.
However, these methods are unable to directly generate novel text: they can only 
provide a compatibility score for a given text--audio input pair.
These models are therefore used for retrieval or keyword detection.

\section{Method}
\label{sec:method}

\begin{figure}
    \centering
    \begin{tikzpicture}[
            font=\scriptsize,
            block/.style={text width=1.25cm, align=center, minimum height=0.75cm, rounded corners},
            block unfrozen/.style={block, fill=red!20, label={[label distance=-10pt]3:\tiny \faUnlock}},
            block frozen/.style={block, fill=blue!20, label={[label distance=-10pt]3:\tiny\faLock}},
            every fit/.style={inner sep=3pt},
        ]
        \newcommand{\mylabel}[1]{{\color{black!50}\tiny\textsf{#1}}}

        \matrix [row sep=0.25cm, column sep=0.75cm] {
                                                     & \node[label=\mylabel{output}] (out) {\it people}; \\
                                                     & \node[block frozen] (head) {head}; \\
                                                     & \node[block frozen] (mlp) {mlp}; \\
    \node[block unfrozen] (proj) {projection}; & \node[block unfrozen] (cross att) {cross attention}; \\
            \node[block frozen] (wav2vec) {wav2vec}; & \node[block frozen] (self att) {self attention}; \\
                                                     & \node[block frozen] (emb) {embedding}; \\[-0.5cm]
            \node[anchor=south, label=below:\mylabel{audio}] (audio) {\includegraphics[height=0.75cm, trim={0 0.5cm 0 0.5cm}, clip]{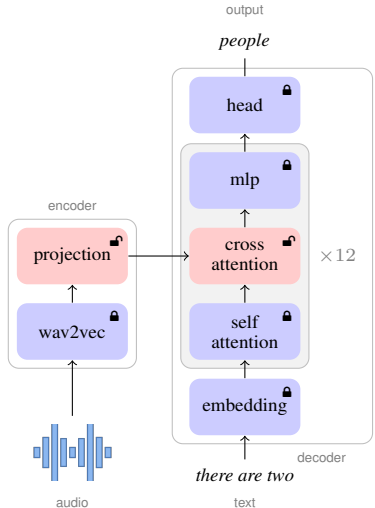}}; &
			\node[anchor=south, label=below:\mylabel{text}]  (text)  {\it there are two}; \\
		};

        \draw[->] (audio) -- (wav2vec);
        \draw[->] (text) -- (emb);
        \draw[->] (wav2vec) -- (proj);
        \draw[->] (proj) -- (cross att);
        \draw[->] (emb) -- (self att);
        \draw[->] (self att) -- (cross att);
        \draw[->] (cross att) -- (mlp);
        \draw[->] (mlp) -- (head);
        \draw (head) -- (out);

        \begin{scope}[on background layer]
            \node[fit=(self att) (cross att) (mlp), fill=gray!10, draw=gray!50, rounded corners, label={[name=layer label]right:{\color{black!50}\scriptsize $\times 12$}}] (layer) {};
            \node[fit=(wav2vec) (proj), draw=gray!50, rounded corners, label={[rotate=0]above:\mylabel{encoder}}] {};
            \node[fit=(emb) (layer) (layer label) (head), draw=gray!50, rounded corners, label=-85:\mylabel{decoder}] {};
        \end{scope}
    \end{tikzpicture}
    \caption{%
        Our audio-to-text model is a transformer that generates text autoregressively conditioned on audio.
        The network consists of learnable (\faUnlock) cross-attention layers interspersed in a frozen~(\faLock) GPT-2 decoder to integrate wav2vec audio features.
    }
    \label{fig:model}
\end{figure}

Our task is speech translation:
given an audio in a foreign low-resource language (Yorùbá) we want to generate a natural language translation in a high-resource language (English).
To this end, we learn an audio-to-text network that generates text autoregressively conditioned on the input audio signal.
We assume that training data consists only of images paired with audio files that describe the contents of the corresponding image.
However, in order to be able to train the audio-to-text network we need audio--text pairs.
We propose to use existing state-of-the-art image captioning systems (such as BLIP~\cite{li2022icml} and GIT~\cite{wang2022tmlr}) to generate captions for the images in the training set.
These text captions paired with the associated audio files %
then serve as data
to train a speech translation model.

While translation is our main task, our method does not make any assumptions on the input and output languages.
If the two languages are the same, for example both the audio files and the image captions are in English,
then our system will perform a type of paraphrasing:
both the input audio and output target text would describe the same semantic information present in the image, but not necessarily using the same words.
This speech paraphrasing task is related but different from the more standard task of automatic speech recognition,
where the output text should contain exactly the same words as the spoken input.

\subsection{Audio-to-text model}

As illustrated in Figure \ref{fig:model}, our audio-to-text model is a transformer model that is composed of two pretrained unimodal models.
The encoder is the wav2vec2 XLS-R 2B model \cite{babu2022interspeech}, which maps the input audio to a sequence of 1920-dimensional embeddings.
The decoder is the GPT-2 model \cite{radford2019gpt}, which generates text in an autoregressive manner.
We couple the encoder and decoder through cross-attention layers, which are inserted after the self-attention layers in each of the twelve GPT-2 blocks.
All parameters of our model are kept fixed, with the exception of the cross-attention layers and a projection layer that maps the audio embeddings (1920D) down to the text space (768D).
Leveraging existing strong pretrained models directly allows for efficient learning in our low-resource setting.
Concretely, our combined transformer has over 2.3B parameters, but only 1.3\% of those (29M) are learnable, making our model lean and more efficient to train.
Our architecture is reminiscent of Flamingo \cite{alayrac2022neurips} or SmallCap \cite{ramos2023cvpr}, but these operate on different modalities (images and text) and have not been employed in our tasks.

We also experimented with an alternative audio-to-text variant:
mapping the audio to a soft prompt to guide the decoding \cite{mokady2021arxiv,manas2023eacl}.
But we found the proposed architecture to work better for our problem.
Another variant that we tried was mapping the audio to image features (instead of text) and use those as input to a frozen image captioner.
But we were not able to make this alternative work as we found it difficult to model the continuous and high-dimensional image embedding space.

\section{Experimental setup}
\label{sec:experimental-setup}

\hspace{\the\parindent}\textbf{Datasets.}
We use two datasets in our experiments:
the Flickr8k Audio Caption Corpus (FACC) \cite{hodosh2013jair,harwath2015asru} for speech paraphrasing and
its Yorùbá counterpart (YFACC) \cite{olaleye2022slt} for speech translation.
FACC is derived from Flickr8k \cite{hodosh2013jair}, which contains 8k images, each annotated with five text captions.
Audio recordings of these captions were subsequently collected by Harwath and Glass~\cite{harwath2015asru}, resulting in 65 hours of data.
YFACC \cite{olaleye2022slt} consists of a subset of the FACC data (one caption per image) that was translated and recorded by a single speaker in Yorùbá;
YFACC totals 13.3 hours.
Although Yorùbá is spoken by roughly 44M people as a first language in West Africa, it is still considered a low-resource language.

\textbf{Metric.}
To evaluate our model, we employ the BLEU metric, a common measure of the similarity of natural texts.
Intuitively, BLEU measures the precision of a hypothesis against a set of reference sentences:
what fraction of the n-grams in a prediction occurs in any of the reference sentences.
We include up to four n-grams, referred to as BLEU-4.
We use the \texttt{sacrebleu} library~\cite{post2018wmt}.
Both the speech translation and speech paraphrasing tasks are evaluated using BLEU.

\begin{figure}
    \centering
    \scriptsize
    \def\myheight{0.75cm}
    \def\mywidth{2.8cm}
	\renewcommand\tabularxcolumn[1]{m{#1}}
    \newcommand\mylabel[1]{\color{gray} \sf \tiny #1}
    \newcolumntype{Y}{>{\centering\arraybackslash}X}
	\renewcommand{\arraystretch}{1.5}
    \begin{tabularx}{\columnwidth}{@{}Y|Y|Y@{}}
        \multicolumn{3}{c}{\includegraphics[width=\mywidth,trim={0 1cm 0 2.5cm},clip]{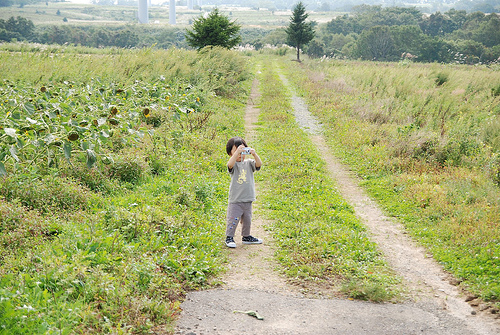}} \\
        \mylabel{beam} & \mylabel{sample} & \mylabel{diverse} \\
        \it A young boy standing on a dirt road next to a field of sunflowers. & \it Many small kids on a path running through a field.                & \it A young boy standing on a dirt road in a field.           \\
        \it A young boy standing on a dirt road next to a field.               & \it A boy is standing and taking photos of his plants on a dirt road. & \it A small child standing in the middle of a dirt road.      \\
        \it A young boy standing on a dirt road in a field.                    & \it A little boy using a camera to look for watermelons.              & \it A child standing on a dirt road in the middle of a field. \\
    \end{tabularx}
    \caption{%
        Sample captions for the image on top using three types of decoding on the GIT image captioning model.
    }
    \label{fig:captions-examples}
\end{figure}

\textbf{Implementation.}
We experiment with three families of image captioning systems (BLIP~\cite{li2022icml}, BLIP2~\cite{li2023icml}, GIT~\cite{wang2022tmlr}) and three types of text decoding techniques (beam search, multinomial sampling, diverse beam search decoding \cite{vijayakumar2016arxiv}).
Some examples are displayed in Figure~\ref{fig:captions-examples}.
For each image we generate five captions using the image captioner.
When training the audio-to-text model we randomly pair each of the five spoken captions with each of the five generated captions.
We use the AdamW optimizer \cite{loshchilov2019iclr} with a learning rate of $1\cdot10^{-4}$, warmed up linearly for 200 steps and then decayed linearly until the end of training.
Training is run for 50 epochs and it takes around six hours on four Tesla T4 GPUs on the YFACC dataset.
We keep the best model as monitored on the development set.
For the translation experiments, we initialize the audio-to-text model from the best model trained on English (FACC),
since this was shown to work better than random initialization~\cite{olaleye2022slt,nortje2023arxiv}.
Our implementation is based on the HuggingFace library~\cite{wolf2019} and is available at
\url{https://github.com/danoneata/strim}.

\section{Experimental results}
\label{sec:experimental-results}

We present our main results and then do a sensitivity analysis to measure the impact of different image captioning methods.

\begin{table*}[!t]
    \def\na{\color{gray}\textsc{n/a}}
    \newcommand\mypm[2]{#1\scriptsize±#2}
    \newcommand\ii[1]{\color{gray} \scriptsize #1}
    \footnotesize
    \centering
    \caption{%
        BLEU scores against the English annotations from the Flickr8k test set (rows 1, 3, 6) or its corresponding Yorùbá subset (rows 2, 4, 5).
        All experiments involving generated captions (rows 3--6) use the GIT image captioning model.
    }
    \begin{tabular}{cll@{\hspace{0.75\tabcolsep}}c@{\hspace{0.75\tabcolsep}}llrrrrr}
        \toprule
        & & \multicolumn{1}{c}{input} & & \multicolumn{2}{c}{targets} & \multicolumn{5}{c}{num. references} \\
        \cmidrule(lr){3-3}
        \cmidrule(lr){5-6}
        \cmidrule(lr){7-11}
        & method & language & & language & decoding & \multicolumn{1}{c}{1} & \multicolumn{1}{c}{2} & \multicolumn{1}{c}{3} & \multicolumn{1}{c}{4} & \multicolumn{1}{c}{5} \\
        \midrule
        \multicolumn{5}{l}{\it Toplines} \\
        \ii{1} & annotator          & \na     &       & \na     & \na         & \mypm{8.32}{0.5} & \mypm{13.95}{1.0} & \mypm{17.84}{0.9} & \mypm{21.59}{0.7} & \na \\
        \ii{2} & translation        & Yorùbá  & $\to$ & English & annotations & \mypm{15.23}{0.0} & \mypm{18.25}{0.3} & \mypm{19.87}{0.4} & \mypm{21.07}{0.3} & \mypm{22.01}{0.0} \\
        \ii{3} & generated captions & \na     &       & English & beam search & \mypm{9.62}{0.9} & \mypm{17.07}{1.0} & \mypm{22.16}{0.8} & \mypm{25.88}{0.6} & \mypm{29.37}{0.6} \\
        \midrule
        \multicolumn{5}{l}{\it Visually grounded speech models} \\
        \ii{4} & translation        & Yorùbá  & $\to$ & English & beam search & \mypm{6.65}{0.0} & \mypm{ 9.37}{0.5} & \mypm{11.32}{0.5} & \mypm{12.72}{0.2} & \mypm{13.71}{0.0} \\
        \ii{5} & translation        & Yorùbá  & $\to$ & English & diverse     & \mypm{6.10}{0.0} & \mypm{ 9.54}{0.6} & \mypm{12.28}{0.9} & \mypm{14.22}{0.4} & \mypm{15.82}{0.0} \\
        \ii{6} & paraphrasing       & English & $\to$ & English & diverse     & \mypm{6.56}{0.5} & \mypm{10.45}{0.8} & \mypm{13.10}{0.7} & \mypm{15.45}{0.4} & \mypm{17.46}{0.9} \\
        \bottomrule
    \end{tabular}
    \label{table:main-results}
\end{table*}

\begin{figure}[!t]
    \centering
    \scriptsize
    \def\myheight{0.75cm}
    \def\mywidth{2.0cm}
	\renewcommand\tabularxcolumn[1]{m{#1}}
    \newcommand\mylabel[1]{\color{gray} \sf \tiny #1}
    \newcolumntype{Y}{>{\centering\arraybackslash}X}
	\renewcommand{\arraystretch}{1.5}
    \begin{tabularx}{\columnwidth}{@{}Y|Y|Y@{}}
        \multicolumn{3}{c}{\mylabel{input audio (Yorùbá)}} \\
		\includegraphics[align=c, height=\myheight, width=\mywidth]{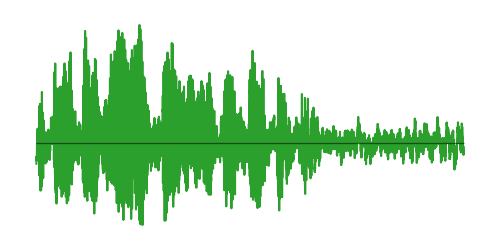} &
		\includegraphics[align=c, height=\myheight, width=\mywidth]{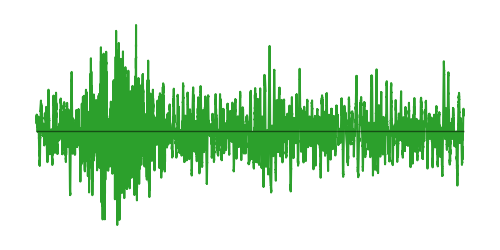} &
		\includegraphics[align=c, height=\myheight, width=\mywidth]{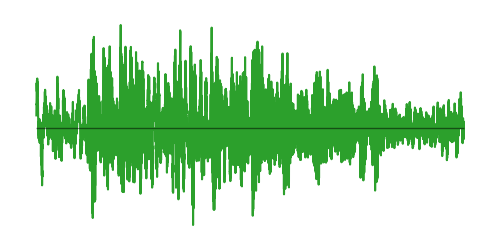} \\
        \multicolumn{3}{c}{\mylabel{groundtruth transcript (Yorùbá)}} \\
		\it \d{O}kùnrin kan dúró leti omi nitosi àpáta.                                      &
		\it Eniyan kan fò ninu af\d{\'e}f\d{\'e}.                                            &
		\it \d{O}m\d{o}kùnrin kan ninu \d{s}okoto penpe pupa ti nmu b\d{\'o}\d{\`o}lù inu agb\d{o}n b\d{\'o}\d{\`o}lù lori pápá. \\
        \multicolumn{3}{c}{\mylabel{groundtruth translation (English)}} \\
		\it A man stands at the edge of the water near the rocks.                &
		\it A snowboarder flies in the air.                                      &
		\it A boy with red shorts is holding a basketball in a basketball court. \\
        \multicolumn{3}{c}{\mylabel{model prediction (English)}} \\
        \it A man standing on a rock with a camera.                              &
        \it A person jumping in the air on a skateboard.                         &
        \it A young boy is playing soccer on a field.                            \\
        \midrule
        \multicolumn{3}{c}{\mylabel{input audio (English)}} \\
		\includegraphics[align=c, height=\myheight, width=\mywidth]{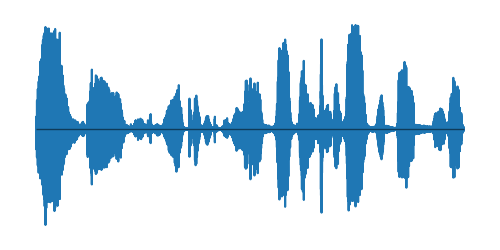} &
		\includegraphics[align=c, height=\myheight, width=\mywidth]{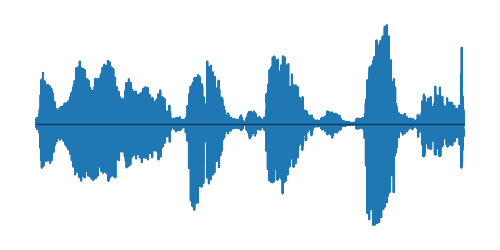} &
		\includegraphics[align=c, height=\myheight, width=\mywidth]{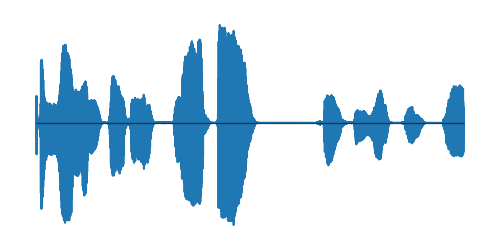} \\
        \multicolumn{3}{c}{\mylabel{groundtruth transcript (English)}} \\
		\it The brown dog is walking through a river surrounded by bushes.                                                         &
		\it Two women in white shirts talking.                                                                                     &
		\it A woman holding a small ball chasing after a small boy.                                                                \\
        \multicolumn{3}{c}{\mylabel{model prediction (English)}} \\
		\it A dog running through the water with its mouth open.                 &
		\it A couple of women standing next to each other.                       &
		\it A young boy holding a baseball bat on a field.                       \\
    \end{tabularx}
    \caption{Examples of Yorùbá-to-English translations (top) and English-to-English paraphrases (bottom) for the visually grounded speech models trained on captions generated by GIT with diverse beam search.}
    \label{fig:qualitative-results}
\end{figure}

\subsection{Main results}
\label{subsec:main-results}

Our main results are given in Table~\ref{table:main-results}.
These are given in terms of the BLEU score (higher is better) against a variable number $n$ of references (captions) for each image,
where $n$ ranges from one to five. With more references, the model gets credit if a predicted n-gram occurs in any of the references; this is reasonable since different people could translate the same sentence differently.
The subset of references is randomly selected from the five captions available for each image.
We repeat each experiment five times and report the mean and two times standard deviation.
The results for three visually grounded speech models (bottom section) are contextualized with three topline systems (top section).

\textbf{Speech translation with images.}
The results for our visually grounded speech translation system are given in rows 4 and 5.
We consider two variants, both using captions generated with the GIT image captioning model, but differing in the type of decoding used: beam search (row 4) or diverse beam search (row 5), as described in Section~\ref{sec:experimental-setup}.
We see that using more diverse captions rather than beam search give slightly better performance.
Performance in absolute terms are modest, but BLEU can be difficult to interpret;
so to give a qualitative indication of performance, the top part of Figure~\ref{fig:qualitative-results} shows sample predictions.
We see that while the audio files are not transcribed verbatim, the predictions do capture the gist of the message being conveyed.
The predictions are valid English sentences, but they tend to be shorter and more direct then the ground truth transcripts.
There are some semantic failures, as 
the model hallucinates the existence of ``camera'' in one example and
mistakes ``snowboard'' for ``skateboard'' and ``basketball'' for ``soccer'' in the other two~cases.

\textbf{Comparison to humans.}
To situate the speech translation results quantitatively, we can compare them to the three topline approaches at the top of Table~\ref{table:main-results}.
The first (row 1) can be seen as human performance on this dataset~\cite{wang2016lrec}: for a given image, we measure how well the caption given by one annotator (hypothesis) matches the captions of others annotators (reference set).
Since each image has five captions, the reference set is limited to a maximum $n$ of four.
The results remain moderate in the absolute: humans reach a BLEU score of only 21.59\% for $n =$~4.
This suggests that even among humans there is a noticeable variance on how they describe the images.
Our best visually grounded speech model, achieving 14.22\% with $n =$ 4, is only 7.36\% behind this topline in absolute BLEU.

\textbf{Comparison to supervised speech translation.} %
Next we consider a direct audio-to-text speech translation model trained on ground truth text annotations (row 2, Table~\ref{table:main-results}).
This model corresponds to the typical speech translation model and we include it to both validate our architecture and put a limit on what is achievable for the visually grounded speech models.
The results are %
even better than the annotator topline for low values of $n$.
For this experiment (as for all those using audio at the input, rows 4--6)
we always include the caption of the input audio caption in the reference set (hence the zero variance when $n =$~1).
So although it might seem surprising at first that this model outperforms the annotators, the model has the advantage of having access to the Yorùbá audio.
As such, this speech translation model can infer the exact words used, while the humans are likely to use different words to cover the semantics.
Comparing this topline to our best visually grounded speech translation approach, we see at $n =$~5 that we are 6.19\% behind in absolute BLEU.

\textbf{How well can we translate with generated captions?}
To answer this question, we consider the performance of the generated image captions (row 3, Table~\ref{table:main-results}), which are used as targets by our speech translation system.
For each image we pick a random image-generated caption as the hypothesis and $n$ annotations as the reference set.
The captions are generated using the GIT model and beam search decoding.
We see that the image captions yield strong results relative to the human annotations, even surpassing the inter-annotator agreement: a BLEU of 25.88\% for $n =$ 4.
This might be caused by the fact that the BLEU metric, being a precision metric, prefers simpler descriptions, which are typically output by image captioning systems.
The performance here are therefore the real upper bound for our visually grounded approach; by comparing our best BLEU of 15.82\% to the 29.37\% at $n =$ 5, we can conclude that the generated captions are not the bottleneck if we want to improve performance.
Rather, other methodological improvements are needed to take advantage of the rich supervision signal present in images.

\textbf{Paraphrasing with images.}
As mentioned in Section~\ref{sec:method}, by using English speech input, we can easily use exactly the same approach as above to do visually grounded speech paraphrasing.
Results for this model is given in row 6 of Table~\ref{table:main-results}.
We see that this speech paraphrasing model comes closer to the annotator and generated caption toplines (rows 1 and 2) than the speech translation models (rows 4 and 5).
But note here that this English--English model is trained and evaluated on the full FACC data, which contains five times more utterances for each image than the YFACC data used for the Yorùbá-to-English speech translation experiments.
The improvement over the Yorùbá--English variants (rows 4 and 5) is therefore presumably 
due to a combination of the larger training dataset and the language match between input and output.
For a qualitative view, sample paraphrases are given in the bottom of Figure~\ref{fig:qualitative-results}.

\subsection{Impact of image captioning}
\label{subsec:sensitivity-analysis}

\begin{figure}
    \includegraphics[width=0.9\columnwidth]{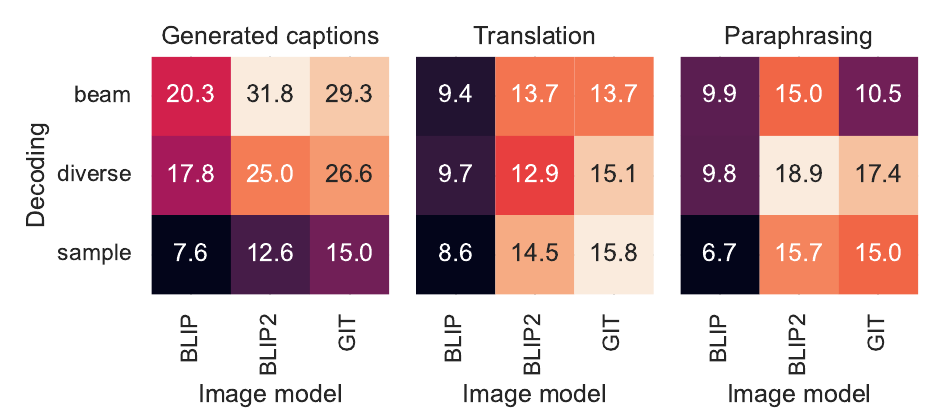}
    \caption{Performance in terms of the BLEU score of the generated captions, speech translation and speech paraphrasing, for all nine combinations of image models and decoding strategies.}
    \label{fig:sensitivity-analysis-image-captioning}
\end{figure}

The image captioning system directly influences the speech translation results
since it provides the targets for the audio-to-text module.
We therefore conduct a sensitivity analysis on three aspects:
the image captioning model, the text decoding strategy, and the number of generated captions.

Concretely, we consider three image captioning models (BLIP, BLIP2, GIT) and three decoding techniques (beam search, diverse beam search, multinomial sampling) and evaluate all nine combinations.
Figure~\ref{fig:sensitivity-analysis-image-captioning} shows the performance of the image captions and of the resulting visually grounded models (both translation and paraphrasing).
The translation models are initialized from the best paraphrasing system.
In terms of the captions performance (Figure~\ref{fig:sensitivity-analysis-image-captioning}-left),
we observe that multinomial sampling performs consistently worse than the other two variants, with beam provides attaining the best results.
The performance across image models is more comparable, but the best results are achieved by the BLIP2 system.

However, these conclusions do not translate for the tasks of interest:
the best variant for translation is the GIT image model with multinomial sampling (Figure~\ref{fig:sensitivity-analysis-image-captioning}-middle), while for paraphrasing it is BLIP2 with diverse beam search (Figure~\ref{fig:sensitivity-analysis-image-captioning}-right).
Notably, multinomial sampling (bottom row) %
now yields the best performance for translation when coupled with the GIT or BLIP2 models.
This suggests that more diverse targets, as illustrated in Figure~\ref{fig:captions-examples}, are important to %
prevent overfitting.

Since diversity is an important factor,
we generated for the translation task a varying number of captions (from one to ten) using GIT captioning with multinomial sampling.
Indeed, when the number of captions is very low (one or two) the performance suffers (9 to 12\% BLEU),
but after three captions, the performance stabilizes at around 15\% BLEU score,
with the maximum of 17.21\% being reached when the number of captions is nine.

\section{Conclusions}

We have shown that it is possible to translate Yorùbá audio to English text using only visual information present in images.
We are able to achieve this by training an audio-to-text model supervised by the text output of an image captioning system.
To build an efficient model, we leverage state-of-the-art components such as wav2vec and GPT-2, and train only a small subset of parameters.
The output predictions convey the semantics of the spoken message in natural language,
but they tend to be simpler and shorter than human translations. %
A limitation of our model is that it tends to hallucinate words, especially when training data is limited.
Future work will explore confidence estimation techniques \cite{tran2022plex} to flag these unreliable predictions.

\section{Acknowledgements}

We thank the anonymous reviewers for their comments and suggestions.
This work was supported in part by European Union’s Horizon research and innovation program under grant agreement No. 101070190 (AI4Trust).

\bibliographystyle{IEEEtran}
\bibliography{myref}

\end{document}